\documentclass[preprint,showpacs,prb]{revtex4}
\begin{document}
\tolerance=5000
\def\be{\begin{equation}}
\def\ee{\end{equation}}
\def\bea{\begin{eqnarray}}
\def\eea{\end{eqnarray}}

\title{ Topological quantization of gravitational fields}
\author{Leonardo Pati\~no}
\email{e.l.patino-jaidar@durham.ac.uk}
\affiliation
   {Department of Mathematical Sciences \\
    University of Durham  \\
    Durham, DH1 3LE, U.K. }
\author{Hernando Quevedo}
\email{quevedo@physics.ucdavis.edu}
\affiliation{Instituto de Ciencias Nucleares\\
Universidad Nacional Aut\'onoma de M\'exico\\
A.P. 70-543,  M\'exico D.F. 04510, M\'exico\\}
\affiliation{Department of Physics \\
University of California \\
Davis, CA 95616}
\date{\today}

\begin{abstract}
We introduce the method of topological quantization for gravitational fields
in a systematic manner. First we show that any vacuum solution of Einstein's
equations can be represented in a principal fiber bundle with a 
connection that takes values in the Lie algebra of the Lorentz group.
This result is generalized to include the case of gauge matter fields
in multiple principal fiber bundles.
We present several examples of gravitational configurations
that include a gravitomagnetic monopole in linearized gravity,
the C-energy of cylindrically symmetric fields, the Reissner-Nordstr\"om
and the Kerr-Newman black holes. As a result of the application
of the topological quantization procedure, in all the analyzed examples
we obtain conditions implying that the parameters entering the metric
in each case satisfy certain discretization relationships.
\end{abstract}

\maketitle

\section{Introduction}

Dirac's original idea \cite{dirac}
of determining the phase acquired by a charge when moving
along a closed path in the field of a magnetic monopole, gave as a result that
the product of the electric charge times the monopole charge is an integer.
Today, this result is known as Dirac's quantization of the electric charge.
In a previous work \cite{lh}, we  introduced a phase-like object which
depends on the field strength so that it can be used to investigate
any field theory based on a connection. In the special case of the 
Levi-Civita connection,
the field strength is given by the Riemann tensor and the phase-like object can
formally be used to investigate the properties of gravitational configurations
that satisfy Einstein's equations. Using only the symmetry properties of the
Riemann tensor and assuming a quite general symmetry property for the 
gravitational
field, we have shown that this phase-like object for vacuum gravitational
configurations behave under rotations either as a bosonic or a fermionic phase.
It was also shown that a certain combination of the eigenvalues of the Riemann
tensor can become ``quantized'' in a fashion similar to that obtained from
Dirac's quantization procedure in the system composed of an electric charge
and a magnetic monopole.

From the geometric point of view \cite{damas}, Dirac's
quantization is interpreted as a consequence
of the existence of a non trivial principal
fiber bundle of $U(1)$ over the sphere $S^2$, with
a $u(1)-$connection, for the system composed of
an electric charge $q$ and a magnetic monopole with magnetic charge $g$.
The Chern numbers associated
with this non trivial fiber bundle turn out to be given as the product
$qg$ which, therefore, becomes quantized.

In this work, we introduce the method
of topological quantization  
which can be applied to any
field configuration whose geometrical structure allows the existence of
a principal fiber bundle. We show that any solution of Einstein's equations 
minimally coupled to any gauge matter field can be represented
geometrically as a principal fiber bundle with spacetime as 
the base space. The structure group (isomorphic to the standard fiber)
follows from the invariance of the orthonormal frame with respect
to Lorentz transformations, in the case of a vacuum solution, 
or with respect to a transformation of the gauge group, in the 
case of a gauge matter field. 
If the bundle turns out to be (globally) non trivial,
the conditions under which
this construction becomes well-defined in all the points where the field
configuration exists, manifest themselves in the transition functions
between different but intersecting open subsets of the base manifold of
the bundle. These conditions on the transition functions turn out to
depend on the parameters which determine the physical structure of the field
configuration. Consequently, the conditions that arise in the construction
of a 
fiber bundle lead to  conditions on the physical parameters which, in turn,
implies that a particular combination of those parameters can take only
{\it discrete} values. This discretization can  be derived also from the
topological invariants of the corresponding non trivial fiber bundle.
This is what we call the {\it quantization conditions}
for a given field configuration. Furthermore, we will see that
even in the case of a globally trivial principal fiber bundle certain 
quantization conditions may appear as a result of demanding
regularity of the connection. 

In Section  II, we introduce the method of topological quantization in
a systematic manner and briefly discuss the general cases in which 
non trivial quantization conditions may appear.
In Section III we analyze the $so(1,3)$-connection of 
cylindrically symmetric gravitational fields and show
that the corresponding C-energy can take only discrete values.
In Section IV we present the example of a gravitomagnetic 
monopole in linearized Einstein's theory which can be 
investigated by means of a $u(1)$-connection. 
Section V contains  the topological quantization 
with respect to  the electromagnetic connection
of gravitational fields which represent electrovacuum 
black holes.
Section VI is devoted to discussions and remarks about future
investigations.

\section{The method of topological quantization}

Consider a Riemannian manifold $(M, g)$, where $M$
is a 4-dimensional differential manifold and $g$ is a bilinear form, the
metric, on $M$. For the purpose of analyzing field equations 
we choose in $M$ a set local orthonormal 1-forms $e^a$, $a=0,1,2,3$.
The orthonormality condition can be expressed in terms of the local
Minkowski metric as $g(e^a, e^b)=\eta^{ab}$, where $\eta^{ab} =
{\rm diag}(+,-,-,-)$. On a torsion free manifold we can introduce 
a connection 1-form $\omega$ by means of the Cartan  first structure
equation
\be
De^a: = d e^a + \omega^a_{\ b} \wedge e^b =0 \ ,
\label{car1}
\ee
where $d$ is the exterior derivative and $D$ the covariant
exterior derivative. We demand that $\omega$ 
be a metric connection, i.e. locally 
$D\eta_{ab}=d\eta_{ab}+\omega_{ab}+\omega_{ba}=0$, a condition 
which implies the antisymmetry of the connection components. 
Furthermore, we use the Cartan second structure equation
to introduce the curvature 2-form $\Omega$ as
\be
D\omega^a_{\ b}:= \Omega^a_{\ b}=
d\omega^a_{\ b} + \omega^a_{\ c}\wedge \omega^c_{\ b} \ ,
\label{car2}
\ee
whose components in terms of the local orthonormal frame,  
$\Omega ^a _{\ b}= (1/2)R^a_{\ bcd} e^c\wedge e^d$, 
determine the Riemann curvature tensor $R^a_{\ bcd}$.
Einstein's gravity theory follows from the variation, with
respect to the orthonormal frame $e^a$ of the 
action
\be
S = -{1\over 32\pi G}\int_M \Omega^{ab}\wedge e^c \wedge e^d \epsilon_{abcd} 
+ \int_M {\cal L}_m \ ,
\label{action}
\ee
where $\epsilon_{0123} = 1$ and ${\cal L}_m$ is the matter Lagrangian which 
depends on $e$, $\omega$, and the matter fields. 
The field equations are 
$\Omega^{ab} \wedge e^c \epsilon_{abcd} = - 16\pi G T_c$,
where $T_c$ is the energy-momentum 3-form which follows from the 
variation of the matter action.     
  
The advantage of using a local orthonormal frame $e$ is that 
the gauge character (at the level of the connection and curvature) 
of Einstein's theory becomes more plausible (see, for instance, 
\cite{damas,gockeler} for a more detailed discussion). Indeed, the 
diffeomorphism invariance of the theory is now reduced to 
the invariance with respect to the Lorentz group $SO(1,3)$.
The change to a different frame $e'=\Lambda  e$ is represented by means
of a matrix $\Lambda  \in SO(1,3)$. The connection 1-form and
the curvature 2-form take values in the corresponding Lorentz
algebra $so(1,3)$, and under a change of frame they transform 
as 
\be
\omega' = \Lambda  \omega \Lambda^{-1} + \Lambda  d \Lambda ^{-1} \ , 
\qquad \Omega' = \Lambda \Omega \Lambda^{-1} \ ,
\label{trans1}
\ee
respectively. It is in this sense that Einstein's theory can be 
considered as a gauge theory with respect to the Lorentz group.
However, it is at the level of the action that Einstein's theory
tremendously differs from pure Yang-Mills gauge theories. 

Let us now consider the matter action. As we have mentioned before
the matter Lagrangian can depend on the frame $e$, the connection 
$\omega$, and the matter fields. We will assume that the matter
fields are gauge fields, that is,   there exists
a connection 1-form $A$, with values in the Lie algebra of a Lie 
group $G$, which generates the gauge field strength $F$ in the standard
manner: $F = d A + A \wedge A$. Under a gauge transformation 
$\gamma \in G$, these quantities behave as
(we assume that the gauge group is a matrix group): 
\be  
A \to A' = \gamma A \gamma^{-1} + \gamma d \gamma^{-1} \ , \qquad
F\to F' = \gamma F \gamma^{-1} \ .
\label{trans2}
\ee

A vacuum spacetime in general relativity is a solution of 
Einstein's vacuum equations, represented by an orthonormal frame 
$e$. Since we assume that the compatibility condition between
the local metric and the connection is satisfied, we can use
the connection $\omega$, instead of the orthonormal frame. 
Moreover, if we adopt the Palatini approach, the connection 
$\omega$ can be considered as the ``primary'' variable, whereas
the orthonormal frame $e$ can be derived from the metricity condition.
In the presence of a matter field, one needs additionally 
the ``matter'' connection $A$ which satisfies Einstein's equations,
with the corresponding energy-momentum 3-form, and the matter field 
equations that follow from the variation of the matter action
with respect to the connection $A$. For a particular spacetime 
to be well defined we have to guarantee that
$\omega$ and $A$ are well defined everywhere in $M$.  
This is a non trivial remark for the analysis of 
gravitational fields we want to perform. Indeed, the 
fact that the connection is demanded to be well defined
everywhere in $M$ implies in general certain conditions that we 
will investigate for explicit fields and which are the fundamental
of what we will call topological quantization. 

Let us be more specific. The idea behind the introduction
 of a differentiable manifold as the  underlying geometric 
structure of spacetime is that in this way one can  
guarantee the existence of coordinate sets 
which cover the entire  manifold. For the approach we are using
here, this is equivalent to having a finite number of sets 
of orthonormal frames covering the manifold. 
Let $\{ U_\alpha\}$ be an open covering of $M$, i.e.
$ \bigcup_\alpha \ U_\alpha = M$.  By definition, a differential manifold
of dimension $n$
is equipped with an atlas $(U_\alpha, \phi_\alpha)$, where 
$\phi_\alpha$ is a homeomorphism from $U_\alpha$ onto an open
subset of the Euclidean space {\bf R}$^n$. If we consider two 
arbitrary open subsets $U_i,\ U_j \in \{U_\alpha\}$ such that
$U_i \bigcap U_j \neq \emptyset $, then in the intersection region the map
$\phi_i \circ \phi_j^{-1}$ is a $C^\infty$ homeomorphism of (open subsets of)
{\bf R}$^n$. Let $\tilde \phi _i$ be the map from $U_i$ into the vector
space of 1-forms $\Lambda^1 (U_i,so(1,3))$ that allows us to introduce
the orthonormal frame $e_i$ in $U_i$. Notice
that the index $i $ labels  different sets of orthonormal frames
and does not refer to any specific component of the frame.
 The orthonormal frame $e_j$ 
attached to $U_j$ by means of $\tilde \phi _j$, is related to 
$e_i$ through an $SO(1,3)$ matrix,  $e_i = \Lambda_{ij} e_j$. It is 
clear that in the intersection region the compatibility condition  
$\tilde \phi _i \circ \tilde \phi _j^{-1} = \Lambda_{ij}$ must be 
satisfied.
On $U_i$ we can also introduce a spin connection 1-form 
$\omega_i$ and a curvature 2-form $\Omega_i$, according to 
Eqs.(\ref{car1}) and (\ref{car2}), respectively. 
These are related to the connection and curvature in $U_j$ 
by means of (no summation over repeated indices) 
\be
\omega_i = \Lambda_{ij}  \omega_j \Lambda_{ij} ^{-1} 
+ \Lambda_{ij}  d \Lambda_{ij} ^{-1} \ , 
\qquad \Omega_i = \Lambda_{ij} \Omega_j \Lambda_{ij}^{-1} \ .
\label{trans3}
\ee

Consider now a third open subset $U_k \in \{U_\alpha\}$
such that $U_i \bigcap U_j \bigcap U_k \neq \emptyset$. 
Accordingly, in the intersection 
region we have that 
$\tilde \phi _i \circ \tilde \phi _k^{-1} = \Lambda_{ik}$ and
$\tilde \phi _j \circ \tilde \phi _k^{-1} = \Lambda_{jk}$. Then, it follows
that 
\be
\Lambda_{ij} \Lambda_{jk} = \Lambda_{ik} \ .
\label{cocycle} 
\ee
This allows us to formulate the following:

\noindent
{\bf Theorem 1:} A solution of Einstein's vacuum field equations can be  
represented by a unique 10-dimensional principal fiber bundle $P$ with the spacetime
$M$ as the base space, the Lorentz group as the structure group
(isomorphic to the standard fiber) and a connection with values 
in the Lie algebra of the Lorentz group.

\noindent
{\it Proof:} A standard theorem (the reconstruction theorem) 
in differential geometry \cite{kn,naber} 
states that a fiber bundle is uniquely specified by the base 
space, the standard fiber, a structure group which is effectively 
represented on the fiber and a family of transition functions, with values in 
the structure group, satisfying the cocycle condition. In the case
of a principal fiber bundle $P$, the fiber is isomorphic to the structure group
which is naturally represented on itself by left translations. For 
a solution of the vacuum field equations, given by an orthonormal frame $e$,
we can take the spacetime manifold $M$ described above as the base space. 
The structure group is identified as $SO(1,3)$. The transition functions
are given by the elements $\Lambda_{ij}: U_i \bigcap U_j \to SO(1,3)$ and satisfy
the cocycle condition which is given above in Eq.(\ref{cocycle}). 
Finally, one can show that it is possible to construct
the projection $\pi: P \to M$ \cite{naber}, once  the transition
functions are given. 
This shows that all the elements of a principle fiber bundle exist and they
can be ``glued'' together to form the desired bundle by means of the 
transition functions and the projection $\pi$.

Finally, we have to show that there exists a connection 
\mbox{\boldmath $\omega$} in $P$. By construction, we do have a 
connection 1-form $\omega$ on $M$, which is the connection associated
with the vacuum solution.  We will see that
it determines a unique connection in $P$. To this end, we use the following 
theorem \cite{gockeler} valid for principal fiber bundles.

\noindent
{\bf Theorem 2:} Given an open covering $\{U_\alpha\}$ of $M$, a structure 
Lie group $G$ with Lie algebra ${\bf g}$,  a family
of local ${\bf g}$-valued 1-forms $\omega_i \in \Lambda^1(U_i, {\bf g})$
which fulfill the compatibility condition 
\be
\omega_i = g_{ji}^{-1} \omega_j g_{ji} + g_{ji}^{-1} d g_{ji} \ ,
\label{comp}
\ee
where $g_{ji}: U_i \bigcap U_j \to G$ are elements of $G$, and a set of local
sections $\sigma_i : U_i \to \pi^{-1}(U_i)$ satisfying $\sigma_j = \sigma_i g_{ij}$
on $U_i \bigcap U_j$, then there is a unique connection \mbox{\boldmath $\omega$}
on $P$
such that $\omega_i = \sigma^*_i \mbox{\boldmath $\omega$}$, where $\sigma^*_i$ is the 
pull-back induced by $\sigma_i$.  

In the case we are considering, the structure group is $SO(1,3)$, the family 
of 1-forms $\omega_i$ is determined by the connection $\omega$ defined
on each $U_i\in \{U_\alpha\}$. The 
compatibility condition (\ref{comp}) coincides with the transformation 
property (\ref{trans3}), once $g_{ji}^{-1}$ is identified with $\Lambda_{ij}$.
It remains to show the existence of local sections. 
Since any fiber bundle accepts a local trivialization which can be defined
as $\Psi_i: \pi^{-1}(U_i) \to U_i\times G$, we can introduce a local canonical section
on $U_i$ by transferring back to $\pi^{-1}(U_i)$ the section of $U_i\times G$, i.e.
by defining $\sigma_i: U_i \to \pi^{-1}(U_i)$ by $\sigma(x) = \Psi_i^{-1}(x,e)$, where
$x\in U_i$ and $e=g_{ii}(x)$ is the identity element of $G$. It is then possible to
show \cite{naber} that this canonical section satisfies $\sigma_j = \sigma_i g_{ij}$
as required. Thus, according to Theorem 2, there exists a unique connection 
\mbox{\boldmath $\omega$} on $P$. This ends the proof of Theorem 1.

We have shown that a vacuum solution can naturally be represented as a 
principal fiber bundle. 
In all steps of the proof, the local connection
$\omega_i$ plays an important role and we have assumed that it satisfies
the continuity and differentiability conditions on $U_i$. The question
arises whether this assumption can be realized in concrete examples 
of gravitational configurations. This is exactly the question that we want
to address in this work by constructing explicitly the elements of 
the principal fiber bundles that correspond to given solutions of 
Einstein's equations. This is what we call the method of 
``topological quantization''. This concept have been
used before in the context of diverse monopole configurations 
\cite{frankel}. We will see that in the process of constructing a 
suitable covering $\{U_\alpha\}$ of $M$, certain ``quantization" conditions appear 
that imply restrictions on the parameters entering the components 
of the connection. 

If it turns out that the constructed principal
fiber bundle admits a global section, the bundle is globally trivial and 
a single connection can be defined everywhere on $P$.  
This could happen, for instance, when the base 
space (spacetime manifold) is contractible. We will see 
that  even in this simple case non trivial conditions 
arise from the requirement that the connection is 
regular on all points of the base space. 
More general cases can be obtained from non contractible 
manifolds which are very common in general relativity. 
 Explicit solutions of Einstein's
equations are usually characterized by the existence of singularities,
i.e. regions that can not be described within the formalism of
general relativity. In order to properly describe the spacetime 
manifold we need to ``remove'' those singular regions from the
manifold. This procedure can be used to obtain non contractible
base spaces for which we can expect that non trivial conditions
arise from the application of the method of topological quantization. 
In fact, non contractible base spaces can give rise to 
globally non trivial principal fiber bundles. In this case one
needs more than one open subset to cover the base space and,
consequently, transition functions appear that turn out 
to generate non trivial ``quantization'' conditions.

Thus, topological quantization is closely related to the problem
of determining whether a principal fiber bundle is globally trivial
or not. 
This is a task that involves the relation between the global 
topological structure of the base space and the fiber, an issue
that is used to perform the classification of bundles and is
related to the theory of characteristic classes and topological 
invariants. Consequently, the method of topological quantization
is closely related to the study of the topological structure
of the underlying bundle. 

From the discussion above it follows that one has (at least) two
ways to perform the topological quantization of a given solution
of Einstein's equations.
The first one consists in using Theorem 1 to construct the corresponding
principal fiber bundle $P$ and the connection \mbox{\boldmath $\omega$}. 
Then, one can analyze the topological invariants of the bundle.
The second method consists in constructing explicitly the 
covering $\{U_\alpha\}$ of $M$ and the family of connection 1-forms
$\{\omega_\alpha\}$ on $M$ for the given solution and extracting
from there the quantization conditions. Obviously, both methods must yield
the same results. When analyzing explicit examples, however,
it is not always easy to construct the unknown principal 
fiber bundle, whereas the second method is straightforward because
we know the connection $\omega$ explicitly. For this reason, in  
this work we will apply mainly the second approach for explicit 
calculations. 

All the above discussion involves only the $so(1,3)-$connection 
associated to the  gravitational action. If an additional matter
action is considered, the results can be formulated in the following
form.

\noindent 
{\bf Theorem 3}: A solution of Einstein's field equations 
 coupled
to a matter gauge field can be  
represented by a unique principal fiber bundle with the spacetime
$M$ as the base space, the gauge group as the structure group
(isomorphic to the standard fiber) and a connection with values 
in the Lie algebra of the gauge group.

The proof of this Theorem is similar to that of Theorem 1.
Indeed, for each open subset $U_i\in \{U_\alpha\}$ we can 
calculate the corresponding gauge connection $A_i$ with values
in the Lie algebra of $G$. The proof can then be carried out 
in a similar manner with 
$\omega_i$ replaced by $A_i$, $\Lambda_{ij}$ replaced by
$\gamma_{ij} \in G$, and $g_{ji}^{-1}=\gamma_{ij}$. 
Consequently, in the case of additional matter gauge fields
we can construct on $M$ a principal fiber bundle for each 
additional gauge connection. So we are lead to the concept
of ``multiple'' principal fiber bundles that can be constructed
on the same base space $M$. 
Since the method of topological 
quantization can be applied to each bundle separately, one could
expect different sets of quantization conditions from each bundle.
The compatibility of these sets is an issue  that can be
treated at the level of explicit gravitational configurations.

In the following sections we will apply Theorems 1 and 3 to
different gravitational configurations.

\section{Cylindrically symmetric gravitational fields}

Cylindrically symmetric vacuum gravitational configurations can 
be described
by means of the Einstein-Rosen line element which in an orthormal 
frame can be written as \cite{kramer} 
\be
e^0 = \exp(\gamma-\psi)dt \ , \quad e^1 = \exp(\gamma-\psi) d\rho\ ,
e^2 = \exp(\psi) d z \ , \quad e^3 = \rho\exp(-\psi) d\varphi \ ,
\label{eiro}
\ee
where $t, \ \rho,\ z,$ and $\varphi$ are cylindrical coordinates and the 
functions $\psi$ and $\gamma$ depend on $t$ and $\rho$ only. 
For the sake of simplicity, here we restrict ourselves to the case 
in which the Killing vector fields $\partial_z$ and $\partial_\varphi$
are hypersurface orthogonal. The 
corresponding vacuum field equations can be reduced to a second-order
differential equation for the function $\psi$
\be
\psi'' + {1\over \rho} \psi' - \ddot\psi = 0 \ ,
\label{eqpsi}
\ee
and two first-order differential equations for $\gamma$
\be
\gamma' = \rho (\dot\psi ^2 + {\psi'} ^2) \ , \quad
\dot \gamma = 2 \rho \dot \psi \psi' \ ,
\label{eqgamma}
\ee
where $\dot\psi = \partial\psi/\partial t$, 
$\psi' = \partial\psi/\partial\rho$, etc. The orthonormal 
frame (\ref{eiro}) is defined up to an arbitrary transformation
of the Lie group $SO(1,3)$.  
If we envision the spacetime as the
4-dimensional base manifold and attach a copy of $SO(1,3)$ at each
point of the base manifold, we obtain the 10-dimensional principal
fiber bundle considered in Theorem 1. Using the first  
structure equation (\ref{car1}) it is straightforward to calculate
the components of the connection 1-form $\omega^a_{\ b}$ which 
can be decomposed as $\omega^a_{\ b} = \omega^a_{\ b\,\mu} d x^\mu$.
It is in this decomposition that the endomorphic character of the 
connection (an 1-form with values in the Lie algebra $so(1,3)$) 
becomes plausible \cite{baez}. From Eq.(\ref{eiro}) we obtain
\bea
\omega^0_{\ 1\, t} & = & \gamma' - \psi'  \ , \quad 
\omega^0_{\ 1\, \rho}  =  \dot\gamma - \dot\psi   \ , \nonumber\\
\omega^0_{\ 2\, z} & = & \dot\psi \exp(2\psi-\gamma) \ , \quad 
\omega^1_{\ 2\, z}  =  -\psi' \exp(2\psi-\gamma)   \ , \\
\label{coneiro}
\omega^0_{\ 3\, \varphi} & = & - \rho\dot\psi \exp(-\gamma)  \nonumber \ , \quad 
\omega^1_{\ 3\, \varphi}  = -(1-\rho\psi') \exp(-\gamma)    \ . \nonumber
\eea

As described in the last section, we have to demand the regularity of this connection
as a condition for constructing the corresponding principal fiber bundle.
It is well known \cite{kramer} that solutions to the field equations can 
be generated which are everywhere regular with the symmetry axis $(\rho=0)$
as the only possible hypersurface where curvature singularities may appear.
Let us suppose that the symmetry axis is free of curvature singularities. 
Then, there must exist an atlas where the connection is also regular. 
The field equations (\ref{eqgamma}) implies that at the axis
$\dot\gamma (\rho\to 0) = \dot\gamma _0 =0$ and $\gamma'(\rho\to 0) = \gamma' _0 = 0$, 
if $\dot\psi (\rho\to 0) = \dot \psi _0$ and $\psi'(\rho\to 0) = \psi' _0$
do not diverge. On the other hand, Eq.(\ref{eqpsi}) implies that near the axis 
$\psi'\propto \rho^\alpha$ with $\alpha>1$, i.e., $\psi'_0 = 0$, and,
 consequently, $\ddot \psi (\rho \to 0) \propto \rho^{\alpha -1}$. Then,
$\dot \psi _0$ is at most a constant that can be set equal to zero
by means of a coordinate transformation. 
Thus we have that the regularity condition at the axis implies that
$\dot\gamma _0 = \gamma'_0 = \dot\psi _0 = \psi'_0 =0$. The same result
can be obtained by analyzing the behavior of the curvature Kretschman 
scalar near the axis. Hence, from Eq.(\ref{coneiro}) it follows that at the axis 
$\omega^a_{\ b}\big|_{\rho=0} = \omega^a_{\ b\,\varphi}\big|_{\rho=0} d\varphi$ with 
\be
 \omega^a_{\ b\,\varphi}\big|_{\rho=0} = \exp(-\gamma_0) T_\varphi \ ,
\ee
where $T_\varphi$ is one of the generators of the Lie algebra $so(1,3)$: 
\be 
T_\varphi=
\left(%
\begin{array}{cccc}
  0 & 0 & 0 & 0  \\
  0 & 0 & 0 & -1 \\
  0 & 0 & 0 & 0  \\
  0 & 1 & 0 & 0  \\
\end{array}%
\right) \ .
\ee
If we demand that $\exp(-\gamma_0)$ does not diverge, the components
$\omega^a_{\ b\,\varphi}\big|_{\rho=0}$ are regular, but we still have a 
singularity in the 1-form $d\varphi=(exp(\psi)/\rho) e^0$. 
Therefore, the only possibility to get rid of this singularity is to 
find a gauge transformation such that the new components 
${\omega'} ^a_{\ b\,\varphi}\big|_{\rho=0}$ vanish identically on the axis.
To this end, let us consider the $SO(1,3)$-transformation
\be
\Lambda = \exp( \tilde \varphi T_\varphi)\ , \qquad 
\tilde\varphi = \exp(-\gamma_0)\varphi \ .
\label{gauge1}
\ee
From Eq.(\ref{trans1}) we have that the gauge-transformed 1-form connection
is given by
\be
\omega' = \exp( \tilde \varphi T_\varphi)\, \omega \, \exp( -\tilde \varphi T_\varphi)
-\exp(-\gamma_0) T_\varphi\, d\varphi \ ,
\ee
where 
\be
\exp(\pm \tilde\varphi T_\varphi) = 1_{4\times4} \pm \sin\tilde\varphi\, T_\varphi 
+(1-\cos\tilde\varphi)T_\varphi^2 \ ,
\ee
where $1_{4\times4}$ is the $4\times 4$ unit matrix. 
The explicit calculation of the components can be carried out in a straightforward
manner and leads to 
\bea
{\omega'} ^0_{\ 1\, t} &=&  (\gamma'-\psi')\cos\tilde\varphi  \ ,\quad
{\omega'} ^0_{\ 3\, t} =  (\gamma'-\psi')\sin\tilde\varphi  \ ,\nonumber \\
{\omega'} ^0_{\ 1\, \rho} &=& (\dot\gamma-\dot\psi)\cos\tilde\varphi   \ ,\quad
{\omega'} ^0_{\ 3\, \rho} =  (\dot\gamma-\dot\psi)\sin\tilde\varphi  \ ,\nonumber \\
{\omega'} ^0_{\ 2\, z} &=& \dot\psi \exp(2\psi-\gamma) \ , \quad
{\omega'} ^1_{\ 2\, z} = -\psi' \exp(2\psi-\gamma)\cos\tilde\varphi \ , \\
{\omega'} ^2_{\ 3\, z} &=&  \psi' \exp(2\psi-\gamma) \sin\tilde\varphi \ , \quad
{\omega'} ^0_{\ 1\, \varphi} =  \rho \dot\psi \exp(-\gamma)\sin\tilde\varphi \ , \nonumber \\
{\omega'} ^0_{\ 3\, \varphi} &=& -\rho \dot\psi \exp(-\gamma)\cos\tilde\varphi \ , \
{\omega'} ^1_{\ 3\, \varphi} = \exp(-\gamma_0)[1 -(1-\rho\psi') \exp(\gamma_0-\gamma)]    \ .
\nonumber 
\eea

From the last expression it can easily be seen that the gauge-transformed connection 
vanishes identically on the symmetry axis and no new singularities appear. 
This has been achieved by means of the gauge transformation (\ref{gauge1}) which is
single-valued only if $\exp(-\gamma_0) = n$, where $n$ is an integer. This, in turn,  
implies that the gauge-transformed connection is single-valued.
Consequently, the condition $\exp(-\gamma_0) = n$ needs to be satisfied for the
connection to be well defined. This is an interesting result that can be interpreted
as a ``quantization" of the energy of cylindrically symmetric gravitational fields.
Indeed, the concept of C-energy was introduced by Thorne \cite{thorne} for 
gravitational fields described by the Einstein-Rosen line element (\ref{eiro})
The quantity $E_c = \gamma_0$ has been shown to represent the (normalized) C-energy
density per length unit along the symmetry axis at a given time. In terms of the
``quantization" derived above this means that $E_c = -\ln n$, i.e. the C-energy
is a discrete quantity. The fact that it is a negative quantity is interpreted by
Thorne as an indication of its ``non classical" origin.  
A second expression that can be considered as a (normalized)
C-energy density has been introduced by Thorne as $E_c = 1-\exp(-2\gamma_0)$. 
In this case, the quantization condition leads to $E_c= 1 - n^2$, an expression 
that again indicates the discrete character of the C-energy. 

We have shown that it is possible to define just one single connection 1-form
on the entire Einstein-Rosen spacetime. This means that the base manifold of
the principal fiber bundle can be covered by a single open set $U$ and,
therefore, can be considered as {\bf R}$^4$ which is a 
contractible manifold. This implies that the corresponding 
10-dimensional principal fiber bundle is globally trivial. 
This is so because we have demanded that the gravitational field be
regular at all points of spacetime, including the symmetry axis. 
If, instead, we would allow singularities on the axis, it would be 
necessary to ``remove" the axis from the base manifold. This would
open the possibility of obtaining a non trivial bundle, an issue
that would require an analysis different from the one presented
in this section.

\section{The weak gravitational field}

Consider the following line element in spherical coordinates 
$t,\ r,\ \theta, \ \varphi$:
\be
ds^2 = (1-2\phi) dt^2 -2\chi dt d\varphi - (1+2\phi)[dr^2 
+ r^2( d \theta ^2 + \sin^2\theta\, d\varphi^2) ]\ ,
\label{wf}
\ee
where $\phi$ and $\chi$ are functions of the spatial coordinates. We
assume that $\phi<<1$ and $\chi<<1$ and consider the weak field limit
of Einstein's equations in vacuum. An orthonormal frame appropriate 
for the line element (\ref{wf}) can be written as
\be
e^0 =(1-\phi)dt + \chi d\varphi  \ , \quad e^1 = (1+\phi)dr \ , \quad 
e^2 = (1+\phi) d\theta\ , \quad e^3 =  (1+\phi)r\sin\theta \ ,
\label{ofwf}
\ee
where we have neglected all the second order perturbations in $\psi$
and $\chi$.

It is well known \cite{wald} that
 the weak field equations in the Lorentz gauge can be expressed as 
the Maxwell equations $\tilde A _{\mu,\nu}^{\ \ \ \nu} = 0$ for 
the Maxwell potential 
\be
\tilde A _\mu = -{1\over 4} (4\phi, \chi) \ ,
\label{maxpot}
\ee
which is invariant with respect to the transformation $\tilde A \to
\tilde A ' = \tilde A + df(x^\mu)$, where $f(x^\mu)$ is an arbitrary 
smooth function. 
This indicates that the weak field approximation can be interpreted
as $U(1)$-gauge theory. To see this explicitly, we introduce the 
$u(1)-$connection 1-form $A = -i \tilde A$. Then, the transformation
law (\ref{trans2}) is identically satisfied for the $U(1)$-gauge
transformation $\gamma = \exp(if(x^\mu))$. 

From the Maxwell equations for this case we see that the 
functions $\phi$ and $\chi$ are decoupled. Let us consider the
following special solution
\be
A = i\left[ \phi dt + {g\over 2} (1+\cos\theta) d\varphi \right] \ ,
\label{mon1}
\ee
where $\phi$ satisfies the differential equation $\phi_{,j} ^{\ \ j}=0$
($j=1,2,3)$, and $g$ is a constant. This is the connection defined
on the base manifold. To investigate
the singularities of the connection (\ref{mon1}), it is convenient
to represent it in the orthonormal frame (\ref{ofwf}). Neglecting
all the second order perturbations we obtain
\be
A_1 = i\left[ \phi e^0 +{g\over 2} {(1+\cos\theta)\over r\sin\theta } e^3 \right] \ ,
\ee
where we have introduced the subscript ``1" to identify it as the connection 
on the open subset $U_1$ that will be determined below. 
We can see that there is a first singularity at $r=0$ which, however, is 
a true curvature singularity as can be seen by analyzing the corresponding
curvature. A second singularity is situated at $\theta=0$ which does not
appear at the level of the curvature. We ``eliminate" the true curvature curvature 
singularity by removing the origin $r=0$ from the spacetime. Hence, the
base manifold $M^4$ becomes $M^4=$ {\bf R}$^4 - \{$world line of $0\}$. 
The second singularity
at $\theta=0$, which corresponds to the positive sector, $z_+$, of the axis $z$,
implies that the connection $A_1$ is regular on 
$U_1 = M^4 - \{z_+\}$. 
This apparent singularity 
can be eliminated by means of the gauge transformation $\gamma = \exp(ig\varphi)$
which leads to the new connection (up to first order in the perturbation)
\be
A_2 = i\left[ \phi e^0 + {g\over 2} {(-1+\cos\theta)\over r\sin\theta } e^3 \right] \ .
\ee
In fact, the singularity at $\theta =0$ has been removed, but a new singularity
has appeared at $\theta = \pi$. Consequently, the connection 
$A_2$ is regular only in the open subset 
 $U_2 = M^4 -\{z_-\}$, where  $\{z_-\}$ denotes the negative $z-$axis. 

The subsets $U_1$ and $U_2$ define a covering of the base manifold. 
In the intersection region $U_1\bigcap U_2$ the two connections are
related by means of the transition function $g_{12} = \exp(ig\varphi)$
which is single-valued only if $g=n$, where $n$ is an integer. 
To interpret this result we have to find out the physical significance
of the parameter $g$. This can be done, for instance,
by calculating the multipole moments of this solution \cite{mm}. To do
this, it is necessary to specify the function $\phi$ and we
chose the simple solution $\phi = m/r$, where $m$ is a constant. 
Then, it can be shown that the parameter $m$ represents the 
monopole moment of a mass distribution and $gm$ corresponds to 
the monopole moment of an angular momentum distribution. This 
implies that $g$ represents the gravitomagnetic monopole per 
mass unit which, according the the quantization condition obtained
above, can take only discrete values.

This example reminds us the case of a magnetic monopole  
in electrodynamics. Indeed, we have chosen the function $\chi$ 
in the connection (\ref{mon1}) as the Maxwell potential 
for Dirac's magnetic monopole. The rest of the analysis is 
then carried out in a similar way as in standard electrodynamics,
due to the analogy between the field equations for the
weak field approximation and the Maxwell equations. 
The result obtained here by analyzing the behavior of the 
$u(1)-$connection can be reproduced in terms of the topological
invariants. The corresponding principal fiber bundle 
is a 5-dimensional $U(1)$-bundle for which the Chern invariants
can be calculated. The result is again that the constant $g$
becomes quantized.

\section{The Reissner-Nordstrom black hole}

Let us consider the following orthonormal frame for the 
Reissner-Nordstrom metric:
\be
e^0 = { [(r-r_-)(r-r_+)]^{1/2}\over r} dt \ , \quad
e^1 = {r\over [(r-r_-)(r-r_+)]^{1/2}} dr \ , \quad
e^2 = r d\theta\ , \quad 
e^3=r\sin\theta d\varphi \ ,
\label{rn}
\ee
with 
\be
r_{\pm} = m \pm \sqrt{m^2-e^2}\  ,
\ee
where $m$ is the mass, $e$ is the net electric charge of
the source and the radial values $r_\pm$ correspond to 
the horizons of the Reissner-Nordstrom black hole.
 This is a solution of the Einstein-Maxwell equations
with the potential $\tilde A = - (e/r) dt$. The corresponding 
$u(1)-$connection $A=-i\tilde A$ behaves under a gauge 
transformation as in Eq.(\ref{trans2}). According to the
discussion of Section II and Theorem 3, there exists a 
principal fiber bundle which can be constructed by 
attaching at each point of the spacetime the fiber 
$U(1)$. 

In this section we will explore the conditions 
that have to be satisfied on the base manifold 
for constructing that bundle. To investigate the 
critical points of the connection 1-form we represent
it in the orthonormal frame (\ref{rn}). Then
\be
A= i e [(r-r_-)(r-r_+)]^{-1/2}\, e ^0 \ .
\label{conrn}
\ee
This connection diverges at $r=r_-$ and $r=r_+$,
whereas the corresponding field strength is regular
at those hypersurfaces.   
To remove these singularities, we first apply
the gauge transformation $\gamma_1 = \exp(iet/r_-)$
on (\ref{conrn}) and obtain
\be
A_1 = -i {e\over r_-} \left({r-r_-\over r-r_+}\right)^{1/2} \  e^0 \ ,
\label{conrn1}
\ee
a $u(1)-$connection which is regular at $r=r_-$, but 
diverges at $r=r_+$. On the other hand, we can also
apply the transformation $\gamma_2 = \exp(iet/r_+)$
on (\ref{conrn}). The resulting connection
\be
A_2 = -i {e\over r_+} \left({r-r_+\over r-r_-}\right)^{1/2} \  e^0 
\label{conrn2}
\ee
is regular at $r=r_+$, but diverges at $r=r_-$.
Thus, we have obtained two different connections
with different divergences. Let us choose 
the open subsets $U_1 = (0,r_+)$ and $U_2 = (r_-,\infty)$.
 This set of open subsets covers
the radial coordinate $r$ completely so that the subsets
$U_1\times ${\bf R}$^3$ and $U_2\times${\bf R}$^3$ are a
covering of the base manifold $M^4$. Then, the connections
$A_1$ and $A_2$ are well defined on $U_1$ and $U_2$,
respectively.  In the intersection
region $U_1\bigcap U_2 = (r_-,r_+)$, the connections $A_1$ and $A_2$ 
have to be related by means of the transition function $g_{12}\in
U(1)$ which can easily be calculated as 
\be
g_{12} = \exp\left[ie\left({1\over r_-} - {1\over r_+}\right)\,t\,\right] \ .
\label{tfrn}
\ee  

The important point about this transition function is that it
depends only on the time coordinate $t$ and is defined only on 
the region contained between the horizons $r_-$ and $r_+$. 
On the other hand, it is well known \cite{mtw} that in this 
region the coordinate $t$ is not timelike but {\it spacelike}.
Indeed, one of the interesting aspects of the region $(r_-, r_+)$
is that the coordinates $t$ and $r$ interchange their role:
what was the radial direction becomes timelike, and the 
timelike direction becomes spacelike. Therefore, we are allowed
to consider $t$ as an angle coordinate $0\leq t\leq 2\pi$ inside
the horizons. This is a consistent procedure that can be carried
out explicitly for all black hole vacuum stationary solutions 
\cite{samos,ohm} and can easily be generalized to the 
case of electrovacuum stationary axisymmetric solutions. 
Therefore, if $t$ is a compact and periodic coordinate
inside the horizons, the transition function (\ref{tfrn})
is single-valued only if the coefficient in front of $t$
is an integer, i.e.
\be
e\left({1\over r_-} - {1\over r_+}\right) = {2\over e}
\sqrt{m^2-e^2} = n \ . 
\label{qcrn}
\ee
This represents a relationship between the physical parameters
which describe the black hole. This specific combination 
of $m$ and $e$ can take only discrete values. Notice that
in the special case of an extreme black hole, $e=m$, the 
only allowed value is $n=0$. Moreover, the limiting case 
$e\to 0$ is not allowed. This is due to the fact that
in order to perform this ``quantization'' 
we have used the $u(1)-$electromagnetic connection which
does not exist in the case $e=0$.

Since the spacetime possesses a curvature singularity at $r=0$,
we have to remove the world line of this event from the base space. 
This implies that the corresponding principal fiber bundle
is not globally trivial. This can be seen explicitly by
calculating the topological invariant, which in this case
is given in terms of the Chern-form $c=-(e/r^2)dt\wedge dr$.
The Chern number is obtained by  integrating the Chern-form
inside the horizon. As expected, we get the value of $4\pi n$,
where $n$ is an integer related to the parameters of the
Reissner-Nordstrom black as given in (\ref{qcrn}). This
represents an alternative derivation of the quantization 
condition. To verify that this result is also independent
of the coordinates, we have performed a similar analysis
of the Reissner-Nordstrom metric in Kruskal-like coordinates.
As expected, the quantization condition (\ref{qcrn}) 
appears in a similar manner.

To conclude this section, it is worth mentioning that 
a similar analysis can be performed for the  
Kerr-Newman black hole. It turns out that in this case
it is necessary to introduce again two open subsets in order
to cover the entire spacetime manifold. The transition 
function is defined in the region contained between the
horizons, where the coordinate $t$ is spacelike, and the 
quantization condition can be written as
\be
{2e^3\sqrt{m^2-a^2-e^2}\over e^4 + 4  a^2m^2 } = n \ ,
\label{qckn}
\ee
where $a$ represents the angular moment per unit mass of 
the black hole. As expected, in the limiting case $a=0$ 
we recover the expression (\ref{qcrn}) for the 
Reissner-Nordstrom black hole.

\section{conclusions}

In this paper we have developed the method of topological quantization
for gravitational field configurations. First, we have shown
that for any vacuum solution of Einstein's field equations
there exists a natural unique principal fiber bundle with an 
$so(1,3)-$connection. If the gravitational field is minimally 
coupled to a gauge matter field, there exists also a principal
fiber bundle with a matter connection. 

This procedure has been carried out explicitly for the gravitational
configurations described by cylindrically symmetric spacetimes,
the gravitomagnetic monopole in linearized gravity and electrovacuum
black holes. In all the cases we have analyzed, the result of the 
topological quantization is a relationship that indicates the
discretization of the parameters entering the corresponding metrics.
We have shown that the quantization conditions arise as the 
result of demanding a regular behavior of the connection 
on the base manifold. Quantization conditions can appear in
globally trivial and non trivial principal fiber bundles. In the
latter case, equivalent results can be obtained from the analysis
of the corresponding topological invariants.

In this work, we do not analyze the physical significance
of the resulting discretization. In particular, it would be interesting
to perform the topological quantization of black holes with respect
to the $so(1,3)-$connection which would complement the result 
of the $u(1)-$connection analyzed here. Preliminary calculations
show that the complete quantization of black holes metrics
leads to a discretization of the horizon area. This task 
is currently under investigation \cite{bhtq}.

Moreover, in all the examples analyzed in this work we have
restricted ourselves to the investigation of the regularity
conditions of the connection on the base manifolds. 
Nevertheless, Theorems 1 and 3 show that there exists an 
additional connection on the bundle which reduces to the
connection on the base manifold, when projected by means
of the pull-back of local trivializations. It would be
interesting to construct explicitly the connection on 
the bundle and investigate its properties. 

Finally, we should mention that although the term ``topological
quantization" could be very suggestive, it is by no means
a procedure that pretends to compete with already existing
and well-developed procedures like canonical quantization.
Nevertheless, it is interesting to see that the mere existence
of relatively simple geometric structures in gravitational 
field configurations leads to a discretization of 
physical parameters, a property that is usually associated
with quantization. A much more detailed and deep investigation
is necessary in order to establish if topological quantization
could be an alternative method to obtain at least partial 
``quantum" information from a physical system.

\section*{Acknowledgments}

L.P. would like to thank C. V. Johnson for 
encouragement and support.
This work was in part supported by
DGAPA-UNAM grant IN112401, CONACyT grant
36581-E, and US DOE grant DE-FG03-91ER40674.
H.Q. thanks UC MEXUS (Sabbatical Fellowship Program)
for support.


\begin{thebibliography}{99}
\bibitem{dirac} P.A.M. Dirac, \textit{Proc. Roy. Soc.} (London) 
\textbf{A133}, 60 (1931).
\bibitem{lh} L. Pati\~no and H. Quevedo, {\it Mod. Phys. Lett. A}
{\bf 18}, 1331 (2003). 
\bibitem{damas} Y. Choquet-Bruhat, C. DeWitt-Morette, and M. Dillard-Bleick,
\textit{Analysis, Manifolds and Physics}, (Elsevier Science Publishers,
Amsterdam, 1982).
\bibitem{gockeler} M. G\"{o}ckeler and T. Sch\"{u}cker, 
\textit{Differential geometry, gauge theories, and gravity}, 
(Cambridge University Press, Cambridge, UK, 1987).
\bibitem{kn} S. Kobayashi and K. Nomizu {\it Foundations of Differential Geometry}
(Wiley Publishers, New York, 1963).
\bibitem{naber} G. L. Naber, {\it Topology, Geometry, and Gauge Fields}
(Springer Verlag, New York, 1997). 
\bibitem{frankel} T. Frankel, {\it  The Geometry of Physics: An Introduction} 
(Cambridge University Press, Cambridge, UK, 1997).
\bibitem{baez}J. Baez and  J. P. Muniain, 
\textit{Gauge Fields, Knots and Gravity}, (World Scientific Publishing, Singapore, 1994).
\bibitem{kramer} D. Kramer, H. Stephani, M. MacCallum, and E. Herlt 
 {\it  Exact Solutions of Einstein's Field Equations} (Cambridge University Press,
 Cambridge, UK, 1980).
\bibitem{thorne} K. S. Thorne, Phys. Rev. {\bf 138}, 251 (1965).
\bibitem{wald} R. M. Wald {\it General Relativity} (The University of Chicago
Press, Chicago, 1984).
\bibitem{mm} H. Quevedo, Phys. Rev. D {\bf 39}, 2904  (1989).
\bibitem{mtw} C. W. Misner, K. S. Thorne, and J. A. Wheeler, {\it Gravitation}
(W. H. Freeman, San Francisco, 1973). 
\bibitem{samos} H. Quevedo and M. Ryan, in {\it Mathematical and Quantum
Aspects of Relativity and 
Cosmology}, eds. S. Cotsakis and G.W. Gibbons (Springer Verlag,
Berlin, 2000).
\bibitem{ohm} O. Obregon, H. Quevedo and M.P. Ryan, Phys. Rev. D {\bf
65}, 024022 (2001); arXiv: gr-qc/0404003. 
\bibitem{bhtq} L. Pati\~no and H. Quevedo, {\it Topological quantization 
of black holes}, In preparation.

\end{thebibliography}
\end{document}